\documentclass[aps,preprint,showpacs,showkeys]{revtex4}
\usepackage{amsmath}
\usepackage{bm}
\usepackage{graphicx}

\renewcommand{\Im}{ {\rm Im\,}}
\renewcommand{\Re}{ {\rm Re\,}}

\begin{document}
\setcounter{page}{1}
\title[]{Local field enhancement in star-like sets of plasmon nanoparticles}

\author{Vladimir  \surname{Poponin}}
\email{ v.poponin@att.net}
\affiliation{Nanophotonics Biosciences Inc., 1801 Bush Street, San
Francisco, CA 94109, USA}
\author{Alexander \surname{Ignatov}}
\affiliation{ General Physics Institute, 38 Vavilova str., 119991
Moscow, Russia}

\date{\today}

\begin{abstract}
We investigate the critical points of the near field intensity in
the vicinity of dielectric bodies illuminated by the incident
light. It is shown that in the electrostatic approximation there
are no local maxima of the intensity outside dielectric surfaces
and the only possible critical points are either local minima or
saddle points. Using the boundary charge method we investigate
numerically the field distribution around star-like sets of
prolate spheroids. The field enhancement is shown to achieve a
value of several hundreds at critical points outside the surfaces
of spheroids.
\end{abstract}



\maketitle

\section{Introduction}

Progress in near-field optics is caused by development of both
experimental technique and mathematical methods of calculations of
the near-field structures. Near-field microscopy has been
successfully used to overcome the diffraction limit and to achieve
subwavelength resolution imaging of various surface structures
\cite{nato} . Of particular interest is a phenomenon of
localization and enhancement of the EM near field by metallic
nanoparticles and clusters \cite{kreibig}. For the explanation of
the phenomena of the Surface Enhanced Raman Scattering of
particular interest are large local fields induced in close
proximity or at the surfaces of nanoparticles \cite
{michaels,jiang}. Mechanism of extraordinary Raman enhancement is
still not completely understood. However, according to
electromagnetic mechanism fields localized on surfaces of
nanoparticles are responsible for extraordinary amplification of
the Raman scattering. According to electromagnetic mechanism
\cite{Mos1,wokaun,schatz} the cross-section of Raman scattering is
proportional to the fourth power of the local field. Therefore,
even moderate enhancement of the local electromagnetic field
results in the extraordinary amplification of the Raman
scattering. Enhancement of the Raman cross section up to ten power
fourteen was reported in literature thus allowing for single
molecule Raman spectroscopy and detection
\cite{c1,michaels,nie,xu}.

Exact solutions to Maxwell's equations are known only for special geometries
such as spheres, spheroids, or infinite cylinder, so approximate methods are
in general required.
Comprehensive review of methods of analysis used in near field optics theory
may be found in review papers and in monographs \cite
{girard,kawata,ohtsu98,ohtsu04}.
Expansion of solution on multipolar eigenfunctions has been applied to many
near-field optical problems in early studies \cite{novotny94}.
 Several approximate methods have been used for calculation of spatial distribution
 of near-field induced by
external electromagnetic wave around nanoparticles of arbitrary
geometrical shape in electrostatic approximation, which is
applicable when  size of nanoparticles is much less than
wavelength of EM field. Among them most popular are
discrete-dipole approximation (DDA) method or coupled dipole
approximation (CDA) method \cite{draine94,kelly03}. Weak point of
DDA or CDA methods is that they can not be applied for particles
with large (exceeding 2) refractive index  because of its low
accuracy and can not describe accurately fine structure with large
field gradients.

Another approximate method commonly used for calculating distribution and
enhancement of local fields is finite difference time domain (FDTD) method
\cite{FDTD}. This method was applied to calculate local field distribution
in a junction between two silver nanoparticles \cite{futumata}. This method
also has disadvantage that it can not describe accurately large field
gradients.

For our purposes we found out that most suitable is Boundary Chage
Method (BCM) when Maxwell's equations are solved numerically using
integral equation formulation \cite{Abajo,xu00}.
Below we will show that by using this method one can more accurately
calculate fine structure of local field distribution in space between
plasmon nanoparticles and on surfaces.

\begin{figure}
\includegraphics{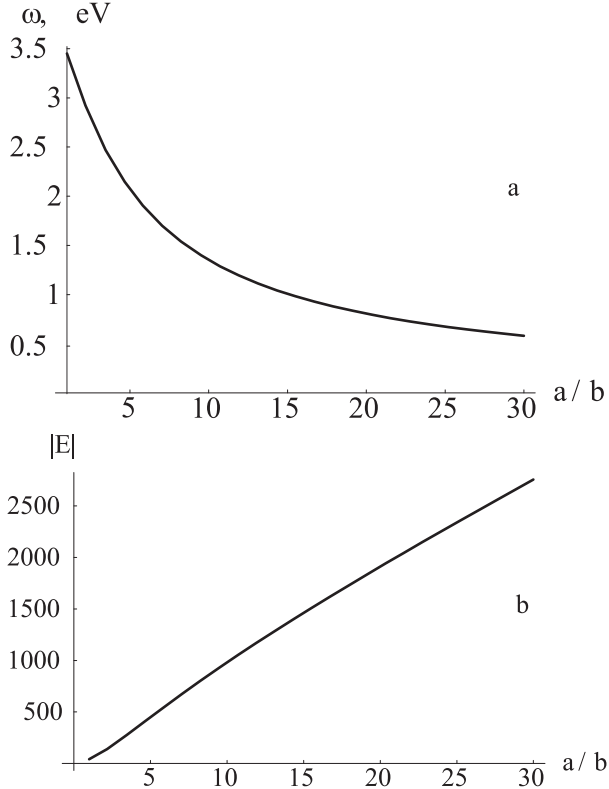}
\caption{The plasmon frequency (a) and the maximal field enhancement (b)
versus the aspect ratio of a prolate spheroid. }
\label{fig1}
\end{figure}

When metallic or dielectric body is exposed to the external
electric field, the maximum field enhancement is achieved at the
points of its surface with maximal curvature. Further field
amplification is obtained by tuning the frequency of the external
field towards the frequency of an appropriate surface plasmon
oscillation. This is illustrated by Fig.~\ref{fig1}, where the
dependence of the frequency of the dipole plasmon resonance (a)
and the maximal field enhancement (b) on the aspect ratio $a/b$
($a>b$) of a single prolate spheroid is depicted using the
well-known solution of the Poisson equation in elliptic
coordinates \cite{mf}. Here the external field is parallel to the
large axis of the spheroid. In plotting Fig.~\ref{fig1}, a simple
approximation for the dielectric function of silver, $\varepsilon
(\omega )=6.4-100/\omega (\omega +0.06i)$, where the frequency,
$\omega $, is measured in eV, was used \cite{kreibig}. As is
readily seen from Fig.~\ref {fig1}, the  field enhancement for a
sufficiently long body may achieve several thousands. Also,
increasing aspect ratio of a spheroidal particle results in a
significant red shift of the plasmon resonance.

Combining particles of various size and shape one may attempt to construct a
nanolens with the region of the maximum field intensity (a hot spot or a
nanofocus) located in a gap between nanoparticles apart from their surfaces.
Recent computations implementing the multiple spectral expansion method \cite
{Li} demonstrated the emergence of such a nanofocus in a system of
self-similar chain of metallic spheres of different size. Such a possibility
is very attractive since it allows for remote detection without the
mechanical contact between a molecule and a nanoparticle.

The main purpose of the present  paper is to investigate the field
distribution near hot spots in sets of plasmon nanoparticles. In
section \ref{secf},  a simple classification of critical points of
the field intensity is given. It is shown that in electrostatic
approximation when one can neglect retardation effects  absolute
maxima of the field intensity are impossible in space outside
surfaces of nanoparticles and the only possible critical points
outside surfaces are either absolute minima or saddle ones.
 The  boundary charge method is explained in
section \ref{secbcm} and  the results of numeric calculations of
the near field distribution produced by symmetric star-like sets
of prolate spheroids are presented in section \ref{secres}.
Finally, in section \ref{sec3} the fine structure of the field
distribution around a self-similar chain composed of metallic
spheres \cite{Li} is investigated and it is shown that maximum
field intensity is achieved on surfaces of nanoparticles. A brief
summary is given in concluding section \ref{seccon}.

\section{ Field distribution near critical points\label{secf}}

When  dimensions of nanoparticles are much less than the
wavelength of an incident  electromagnetic field,  one can neglect
retardation effects and use the electrostatic approximation to
calculate the field distribution. Thus in the near-field region,
the electric field, $\bm{ E}(\bm{ r})=-\nabla \varphi (\bm{ r})$,
satisfies Poisson's equation,

\begin{equation}  \label{p0}
\nabla \varepsilon(\omega , \bm{ r})\nabla \varphi(\bm{ r})=0,
\end{equation}
with the boundary condition
\begin{equation}  \label{bc}
\bm{ E}(\bm{ r}\to\infty)\to \bm{ E}^{in},
\end{equation}
where $\varepsilon(\omega , \bm{ r})$ is the complex dielectric
permittivity and $\bm{ E}^{in}$ is the electric field amplitude of
the incident electromagnetic wave. In traditional electrostatic
problems, when the dielectric permittivity is independent of
$\omega$, both the external field and the potential are
real-valued functions. In application to the near-field optics,
 both the electric field of the incident wave, $\bm{ E}^{in}$, and
 the potential, $\varphi(\bm{ r})$, may be complex-valued due
to the possible elliptical polarization of the incident wave and due to the
energy losses inside nanoparticles.

Prior to discussing the results of numeric solution to Eq.~(\ref{p0}), we
give the qualitative description of hot spots. To be more rigorous, we
define a hot spot as a point where the field intensity distribution has a
maximum in certain directions. In other words, we are interested in the
critical points with zero gradient, $\nabla I(\bm{ r})=0$, where the field
intensity is $I(\bm{ r})=|\bm{ E}(\bm{ r})|^2=|\nabla\varphi(\bm{ r})|^2$.
The question is what kind of critical points are possible apart from the
surface.

To answer this question suppose that the critical point is at the origin, $%
\bm{ r}=0$. In a vicinity of this point, the solution to Poisson equation (%
\ref{p0}) may be written as a sum
\begin{equation}  \label{pexp}
\varphi(\bm{ r})= \varphi_0-\bm{ E}_0 \cdot\bm{ r}+\varphi_2(\bm{ r}%
)+\varphi_3(\bm{ r})+\dots,
\end{equation}
where $\bm{ E}_0$ is the electric field at the origin and $\varphi_n(\bm{ r})
$ are harmonic polynomials of the order $n$. In particular, the quadratic
and cubic parts of the expansion are

\begin{eqnarray}  \label{v2}
\varphi_2(\bm{ r})&=&A_2 \left\{ 3 (\bm{ n}_1\cdot \bm{ r})(\bm{
n}_2 \cdot
\bm{ r})-(\bm{ n}_1 \cdot\bm{ n}_2)r^2 \right\},
\\
\label{v3} \varphi_3(\bm{ r})&=&A_3 \{ 5 (\bm{ m}_1 \cdot\bm{
r})(\bm{ m}_2 \cdot \bm{
r}) (\bm{ m}_3 \cdot \bm{ r}) \\
&-&r^2[ (\bm{ m}_1 \cdot \bm{ r})(\bm{ m}_2 \cdot \bm{ m}_3)+ (\bm{ m}_2
\cdot \bm{ r})(\bm{ m}_1\cdot \bm{ m}_3)+(\bm{ m}_3 \cdot\bm{ r})(\bm{ m}%
_1\cdot \bm{ m}_2)] \}  \notag
\end{eqnarray}
where $A_{2,3}$ are arbitrary constants and $\bm{ n}_{1,2}$ and $\bm{ m}%
_{1,2,3}$ are arbitrary unit vectors. The field intensity near the origin
looks like

\begin{equation}  \label{fe}
I(\bm{ r})=E_0^2- 2 \Re (\bm{ E}_0^\ast \cdot\nabla \varphi_2(\bm{ r}))+
|\nabla\varphi_2(\bm{ r})|^2 -2 \Re (\bm{ E}_0^\ast \cdot\nabla\varphi_3(%
\bm{ r}))+\dots
\end{equation}

Let us consider first the case of linear polarized incident wave;
then $\bm{ E}^{in}$ in Eq.~(\ref{bc}) is a real vector. If the
losses inside dielectric particles are negligible, $\Im
\varepsilon(\omega)=0$, the resulting solution to the Poisson
equation is a real-valued function.
According to Eq.~(\ref{fe}) the condition of zero intensity gradient at $%
\bm{ r}=0$ is $\bm{
E}_0\cdot\nabla\varphi_2\equiv0$. It is a matter of simple algebra to check
that this yields either to $\bm{ E}_0=0$ or to $A_2=0$ in Eq.~(\ref{v2}).
The first possibility, $\bm{ E}_0=0$, corresponds to the absolute minimum of
the field intensity, $I(0)=0$; this field configuration is used in the
well-known Paul traps. The second possibility means that $\varphi_2(\bm{ r}%
)=0$ in Eqs.~(\ref{pexp},\ref{fe}), hence the quadratic part of the
expansion in Eq.~(\ref{fe}) is $-2 \bm{ E}_0 \cdot\nabla\varphi_3(\bm{ r})$.
The latter expression is a harmonic function. According to the well-known
properties of harmonic functions \cite{mf} it cannot take maximal or minimal
values inside its domain. The only possible critical points are the saddle
ones. Thus, with a linear polarization of an incident wave and negligible
losses the field intensity apart from the surface of dielectric bodies may
be either zero at certain points or it may exhibit a saddle point.

In the case of complex-valued fields the analysis is similar but more
tedious. The zero-gradient condition, $\nabla I(\bm{ r})=0$, does not
necessarily means that $\varphi_2(\bm{ r})$ in Eq.~(\ref{pexp}) vanish. In
order to classify the critical points of the field intensity we have to
investigate the eigenvalues of the quadratic form $I_2=|\nabla\varphi_2(\bm{
r})|^2 -2 \Re( \bm{ E}_0^\ast \cdot\nabla\varphi_3(\bm{ r}))$ appearing in
Eq.~(\ref{fe}). Since all coefficients and vectors entering in Eqs.(\ref{v2},%
\ref{v3}) are generally complex, the quadratic form $I_2$ depends
on 12 complex numbers. We have studied the signature of $I_2$
using Mathematica 5.0 symbolic software. It was found that three
eigenvalues are never negative all at once, that is, the field
intensity (\ref{fe}) never takes maximal value apart from
dielectric surfaces. The signature of the quadratic form, $I_2$,
may be either $(+,-,-)$, $(+,+,-)$ or $(+,+,+)$. The latter
possibility corresponds to the local minimum of the field
intensity that may now exist even if $\bm{ E}_0\neq0$. Although
such a ``cold spot'' is hardly of use in the near-field optics, it
is worth writing down a corresponding example of the field
distribution:

\begin{equation}  \label{zero}
\varphi(\bm{ r})=-x + 2y + z+ 2 i xy + \frac{1}{8}x^2y - \frac{1}{4}y^3 +
\frac{1}{4}x^2 z + \frac{1}{4}x yz - \frac{1}{4}y^2 z + \frac{5 }{8}y z^2
\end{equation}

Thus the most important conclusion from this analysis is that the field
intensity cannot take maximal values apart from the dielectric surfaces. The
spatial distribution of the intensity may exhibit either saddle points or
minima. Below we demonstrate several numeric examples with both types of
saddle points.

\section{ Boundary charge method\label{secbcm}}

For the case of materially equal dielectric particles the problem is
conveniently reduced to a set of integral equations for the surface charge
density, $\sigma$, induced at the nanoparticles \cite{Abajo} or,
equivalently, for the surface distribution of the normal electric field \cite
{me}. Compared to various direct methods of the numeric solution of Eq.~(\ref
{p0}), the integral equation approach allows to reduce dimensionality of a
problem and to achieve reasonable accuracy at smaller computer cost. Here,
we use the boundary charge method (BCM) based on the integral equations in
the form derived in \cite{Abajo}:

\begin{equation}  \label{int1}
\Lambda(\omega)\sigma(\bm{ s})=-\bm{ n}(\bm{ s})\cdot\bm{E}^{in}(\bm{ s})+
\int ds^\prime F(\bm{s},\bm{s}^\prime)\sigma(\bm{ s}^\prime),
\end{equation}
where
\begin{equation}  \label{lambda}
\Lambda(\omega)=2\pi\frac{1+\varepsilon(\omega)}{1-\varepsilon(\omega)},
\end{equation}
and
\begin{equation}  \label{kern}
F(\bm{ s},\bm{ s}^\prime)=-\frac{\bm{ n}(\bm{ s})\cdot(\bm{ s}-\bm{ s}%
^\prime)}{|\bm{ s}-\bm{ s}^\prime|^3}.
\end{equation}

The integration in Eq.~(\ref{int1}) is carried over the multiply connected
surface of all particles, vectors $\bm{ s}$ and $\bm{
s}^{\prime }$ belong to the surface, and $\bm{ n}(\bm{ s})$ stands for the
surface unit normal directed towards vacuum at $\bm{ s}$. Once Eq.~(\ref
{int1}) is solved and the surface charge distribution, $\sigma (\bm{ s})$,
is found one can calculate the induced electric field at the arbitrary point
outside the surface using a simple formula

\begin{equation}  \label{ff}
\bm{E}(\bm{r})=\bm{E}^{in}-\nabla_{\bm{r}}\int ds\,\frac{\sigma(\bm{s})}{|%
\bm{r}-\bm{s}|}.
\end{equation}

Eq.~(\ref{int1}) was approximated by a set of linear equations with the help
of an appropriate triangulation of the surfaces. Our computer facilities
allowed us to use up to 5000 triangulation points. As a test problem, we
investigated the dipolar plasmon resonance for a single prolate spheroid.
Comparing the numeric solution with the well-known analytic expressions (see
fig.~\ref{fig1}), it was found that with the maximum affordable number of
triangulation points the accuracy within 10\% is achieved for the spheroid
aspect ratio not exceeding $6:1$.

\begin{figure}
\centerline{\includegraphics{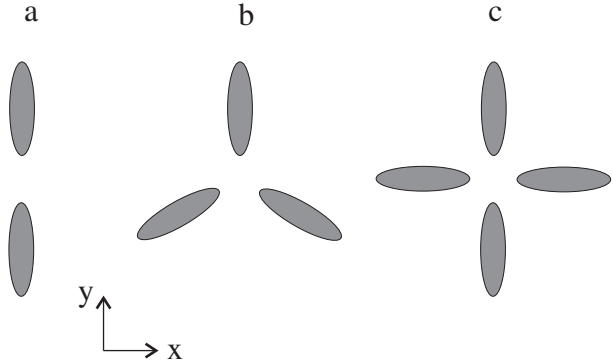}} \caption{Sets of prolate
spheroids used in computations. } \label{fig2}
\end{figure}

\section{ Results of computations for star-like sets of nanoparticles\label{secres}}

Now we turn to the discussion of the numeric solutions to integral equation (%
\ref{int1}). We have investigated the field structure near the sets of
prolate spheroids (Fig.~\ref{fig2}). For the examples discussed below, the
aspect ratio of all spheroids is $4:1$, their centers are at $r=4.5$ from
the origin. The complex dielectric permittivity, $\varepsilon(\omega)$,
corresponds to silver.

\begin{figure}
\centerline{\includegraphics{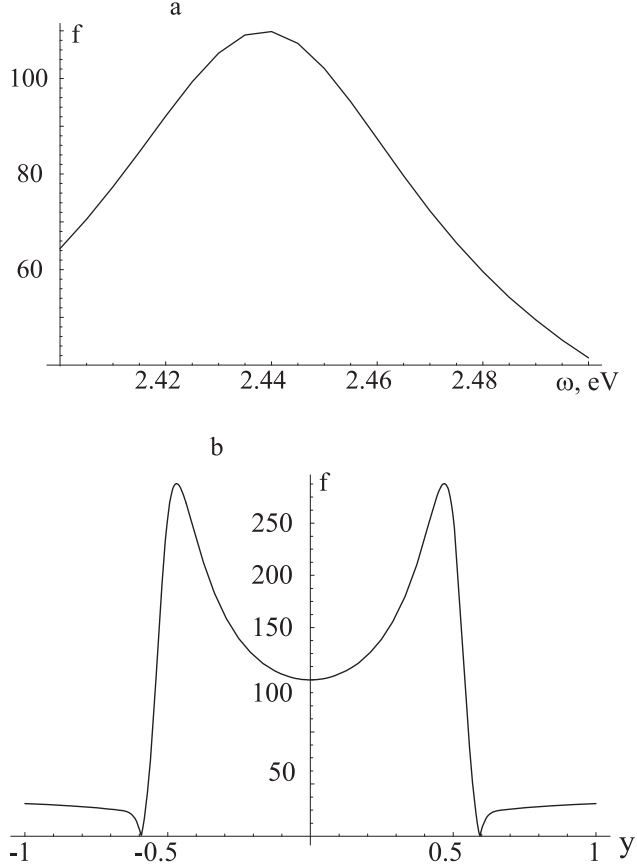}} \caption{(a) --- The field
at the origin versus frequency of the incident wave, (b) the field
distribution along the $y$ axis ($\protect\omega=2.44$ eV) for the
set shown in Fig.~\ref{fig2}a.} \label{fig3}
\end{figure}

Fig.~\ref{fig3} shows the dependence of the field enhancement, $f=|\bm{ E}|/|%
\bm{ E}^{in}|$, at the origin on the frequency of the incident wave for the
configuration depicted in Fig.~\ref{fig2}a. The incident wave propagating
along the $z$ axis is linear polarized with the electric field vector along
the $y$ axis, $\bm{ E}^{in}=(0,1,0)$. The field distribution along $y$ axis
is also shown. The electric field here is maximal at the surfaces of
spheroids. At the origin, there is the maximum in $zx$ plane and the minimum
in $y$ direction. The signature of the saddle point at $\bm{ r}=0$ is $%
(-,+,-)$. Notice that the maximum field enhancement for a single $4:1$
prolate spheroid is about $f\approx 300$ (fig.~\ref{fig1}), that is, the
maximum field at the origin is approximately half as much. With the circular
polarization of the incident wave, there are just minor changes in the field
structure.

\begin{figure}
\includegraphics{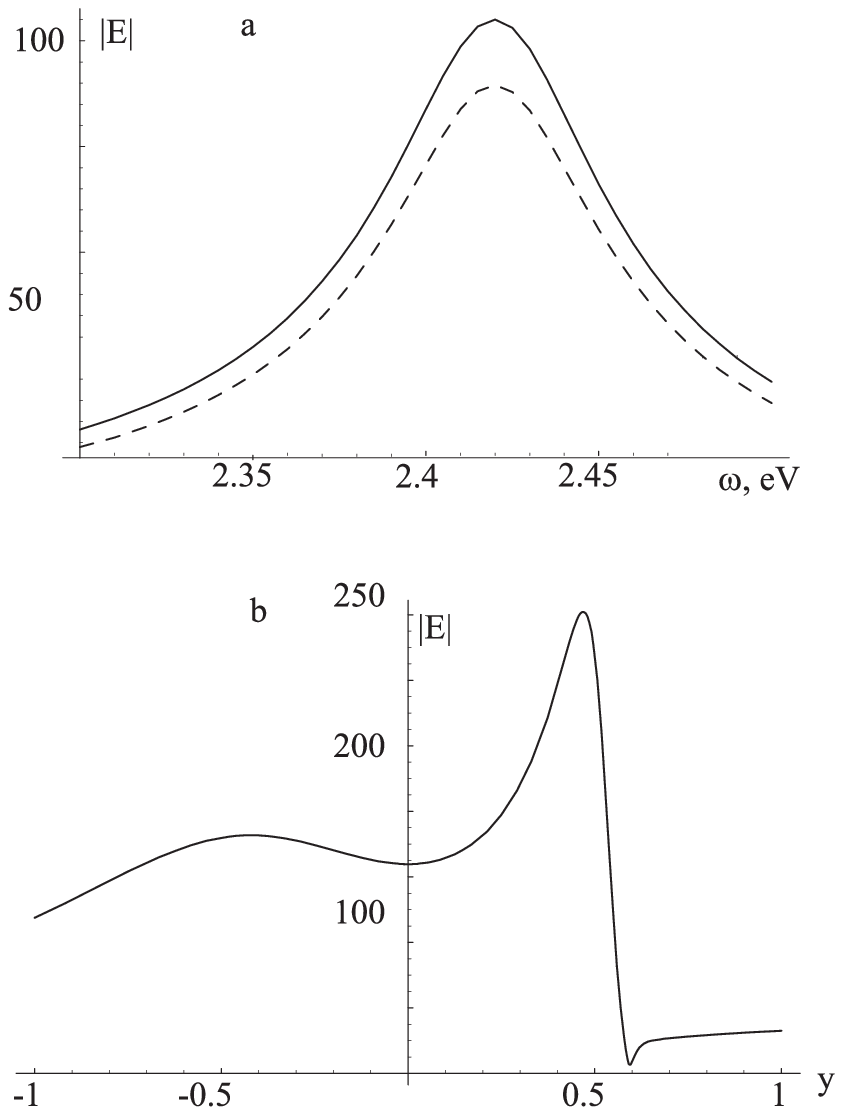}
\caption{a --- Field amplitude versus frequency for the circular
polarization of the incident wave and configuration depicted in
Fig.~\ref
{fig2}b. Solid line --- $\bm{ r}=(0,-0.45,0)$, dashed line --- $\bm{r}%
=(0,-0.45,0)$. b --- Field distribution along $y$ axis. }
\label{fig4}
\end{figure}

\begin{figure}
\includegraphics{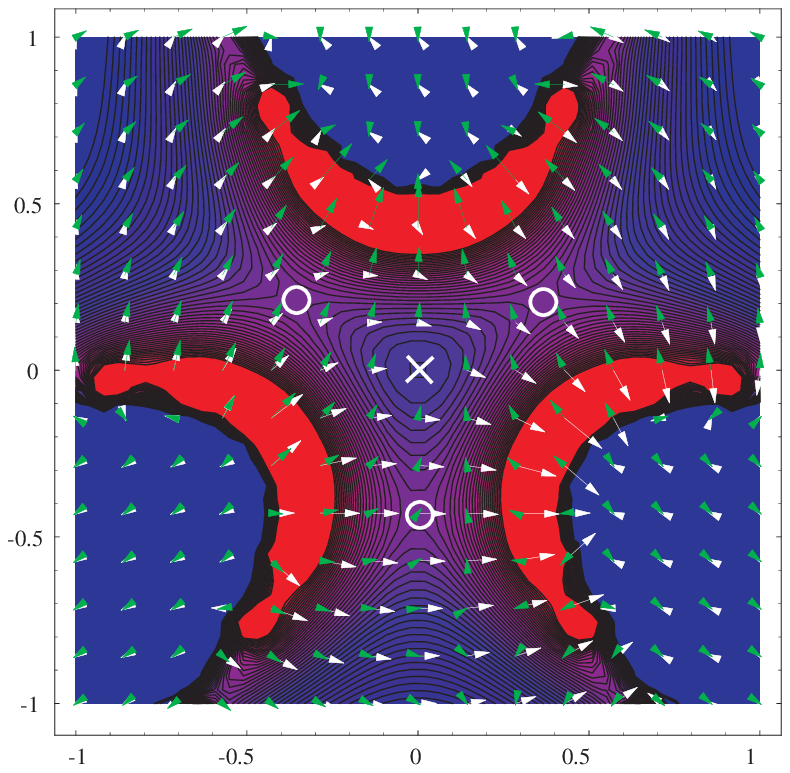}
\caption{Field distribution in $xy$ plane around three spheroids
(Fig.~\ref {fig2}b) with circular polarization of the incident
wave. $\times$ corresponds to $(+,+,-)$ saddle point, $\bigcirc$
corresponds to $(+,-,-)$ saddle points. Green arrows --- $\Re \bm{
E}$, black arrows --- $\Im \bm{ E} $. } \label{fig5}
\end{figure}

More complicated behavior is observed for the star composed of three
spheroids (Fig.~\ref{fig2}b). Fig.~\ref{fig4}a shows the field versus the
frequency for the circular polarization of the incident wave, $\bm{ E}%
^{in}=(1,i,0)/\sqrt{2}$. The plasmon resonance frequency (2.42 eV) in this
case is a little red-shifted compared to two spheroids (2.44 eV). The field
structure, however, is entirely different. The field distribution along the $%
y$ axis is depicted in Fig.~\ref{fig4}b. The two-dimensional contour plot of
the field intensity is shown in Fig.~\ref{fig5}, where the the electric
field vectors are also plotted.

Instead of a single saddle point between two spheroids now there are four
critical points. The minimum in the $xy$ plane (marked with the $\times$
sign in Fig.~\ref{fig5}) is situated at the origin. Three others ($\bigcirc$
) are maximums in the radial direction and minimums in the azimuthal
direction. The field enhancement at these points is about $|\bm{ E} |\approx
150$. Fig.~\ref{fig5} also shows the real and imaginary parts of the
electric field vector. As is readily seen, the ellipticity of the wave is
essentially nonuniform, i.e., this set of spheroids acts as a polarizer.

\begin{figure}
\includegraphics{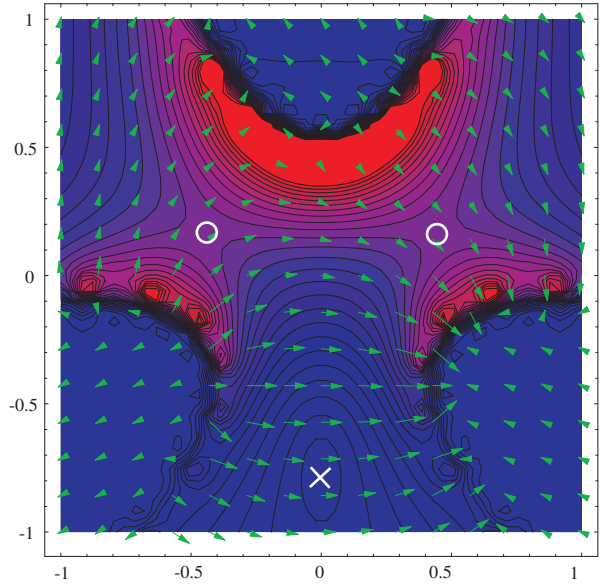}
\caption{Same as in Fig.~\ref{fig5} but with linear polarization
of the incident wave. } \label{fig6}
\end{figure}

In contrast with two spheroids, the field structure in the 3-star is
sensitive to the polarization of the incident wave. Fig.~\ref{fig6} shows
the intensity distribution and the electric field for the linear
polarization of the incident wave, $\bm{
E}^{in}=(0,1,0)$. Now there are only two $(+,-,-)$ saddle points with the
field enhancement $|\bm{ E }|\approx 130$. Besides, the $(+,+,-)$ point
moves to the lower part of the figure.

\begin{figure}
\includegraphics{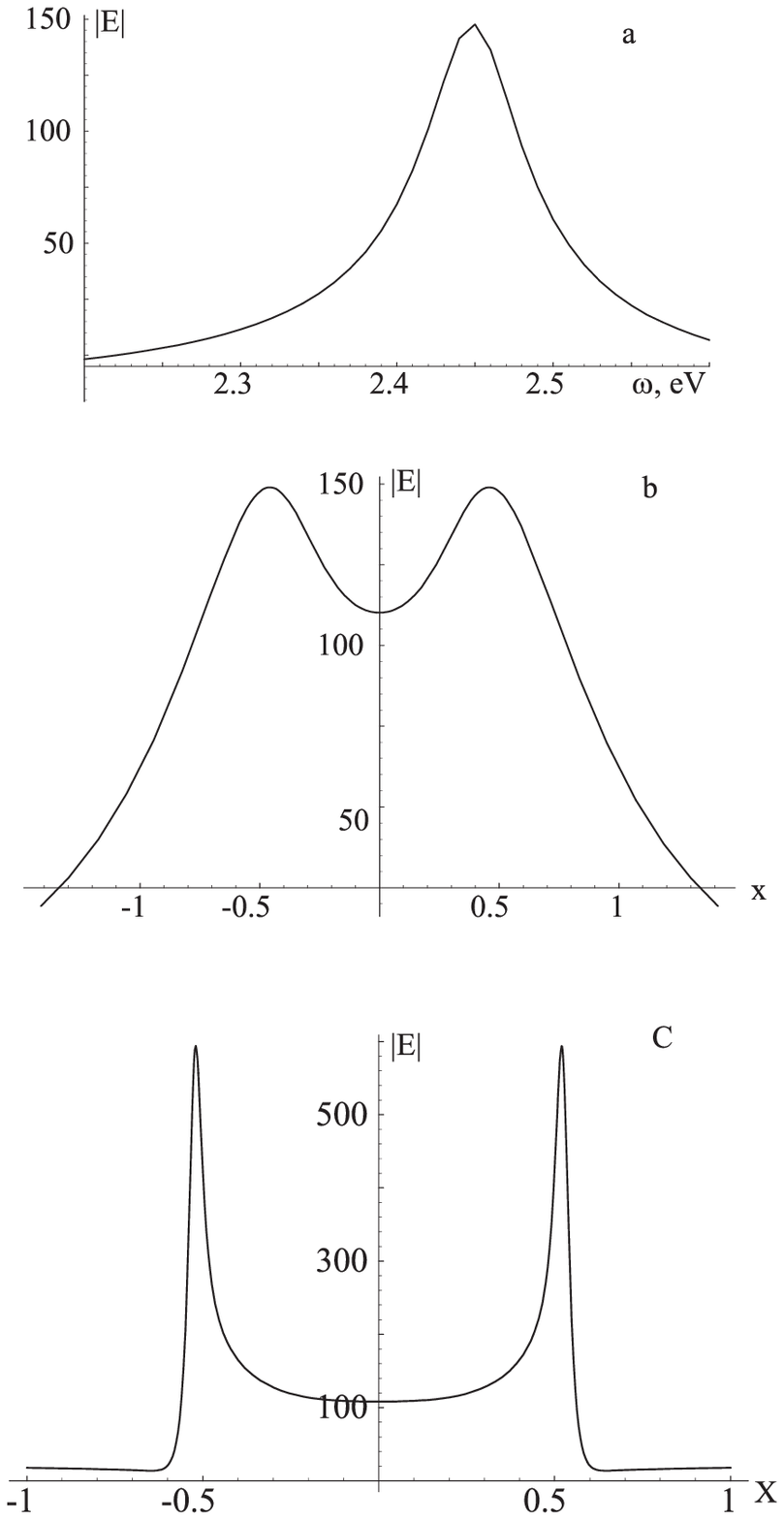}
\caption{a --- Field amplitude versus frequency for the circular
polarization of the incident wave for the set shown in
fig.~\ref{fig2}c. b
--- Field distribution along the $y=x$ line. c --- Field distribution along
the $x$ axis. }
\label{fig7}
\end{figure}

\begin{figure}
\includegraphics{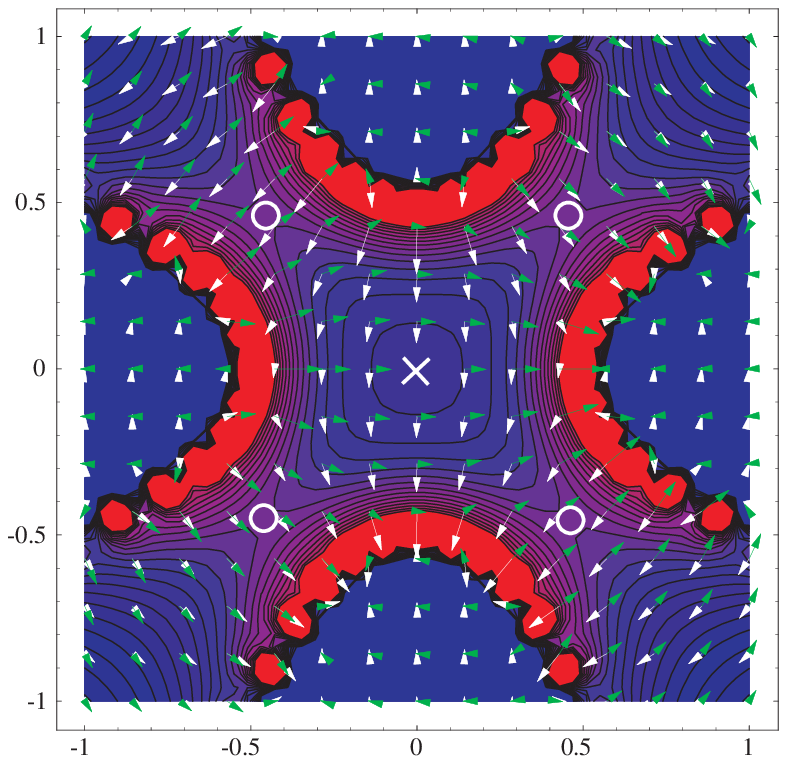}
\caption{Same as in Fig.~\ref{fig5} for the 4-star set
(Fig.~\ref{fig2}c). } \label{fig8}
\end{figure}

Similar behavior was observed for the structures consisting of larger number
of prolate spheroids. Fig.~\ref{fig7}a shows the plasmon resonance for the
circular polarization of the incident wave. The field distribution along a
line coming through the gap between two adjacent spheroids is depicted in
fig.~\ref{fig7}b. The two-dimensional plot (Fig.~\ref{fig8}) demonstrates a
single $(+,+,-)$ saddle point and four $(+,-,-)$ saddle points.

\section{ Field enhancement by a self-similar chain of spheres\label{sec3}}

As we have already mentioned, it was recently proposed to achieve larger
field enhancement implementing self-similar sets of nanoparticles \cite{Li}.
The idea is based on the fact that in the near-field region the field
enhancement is independent of absolute size of particles. Therefore, if
there two spheres of different size separated by a small gap, then the
resulting field enhancement at smaller sphere should be about $f^2$, where $%
f $ is the enhancement factor for a single isolated sphere, which depends on
its material only. Combining $m$ spheres of reducing radii one may expect to
achieve the electric field at the surface of the smallest sphere enhanced by
a factor of $f^m$.

\begin{figure}
\includegraphics{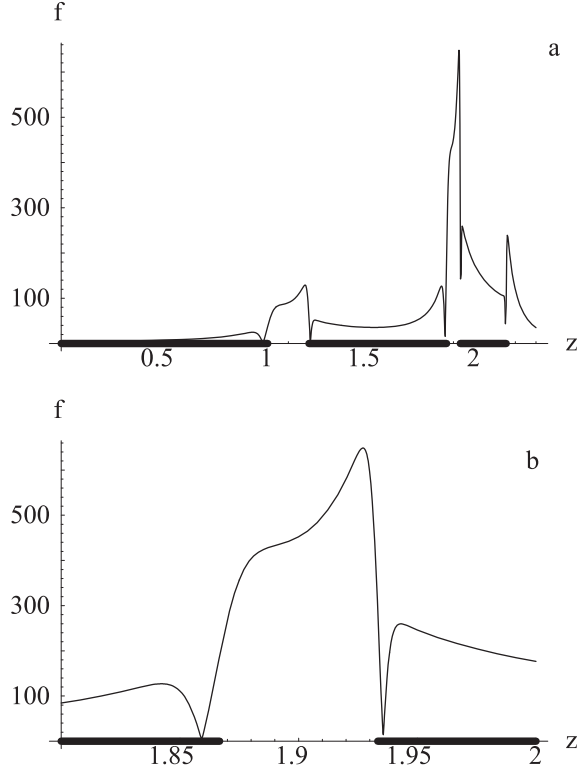}
\caption{Field distribution along a chain composed of three silver spheres, $%
\protect\omega=3.35 eV$. Heavy lines mark the position of the spheres. }
\label{fig9}
\end{figure}

We used BCM to investigate a combination of three silver spheres
studied in \cite{Li}. The radii of spheres are $R=1,\; 1/3,\;
1/9$, their centers are at $z=0,\; 1.53, \; 2.0$. The electric
field of the incident wave is parallel to the $z$ axis. The field
distribution along the $z$ axis is depicted in Fig.~\ref{fig9}. As
is readily seen from the figure, the field at the surface of the
smallest sphere is really enhanced by a factor of 600. However,
the maximal field is achieved rather at the surface of the sphere
then in a gap between spheres (see Fig.~\ref{fig9}b, where the
magnified part of Fig.~\ref{fig9}a is shown).

\section{ Conclusion\label{seccon}}

It is shown that in the nonretarded approximation the absolute
maximum of the field intensity distribution around dielectric
bodies illuminated by an external source is achieved at the
surface. The only possible critical points of the intensity
distribution outside surfaces are either saddle or minimal points.
The computed field distribution around the sets of prolate
spheroids having the aspect ratio factor up to 4  exhibits the
maximum field enhancement on the order of 300 at the points of the
surface with maximal curvature. It is worthy to note, that the
field enhancement provided by a single silver sphere is about 30
\cite{futumata}. However, in star like set of nanoparticles the
field enhancement at the saddle points (''hot spots'') outside the
surfaces is also sufficiently large, $ f\approx 150$. even for the
aspect ratio 4. Since the Raman scattering is proportional to the
fourth power of the electric field, the corresponding Raman
cross-section is enhanced by a factor of $5\;10^{8}$ . Preliminary
computations with the increased aspect ratio and the fixed
distance between spheroids demonstrated, first, the significant
red shift of the plasmon resonance. Second, the field enhancement
at the saddle points is approximately half the enhancement at the
tip of an single spheroid with the same aspect ratio, that means,
that field amplification at central points may achieve a value of
several thousands for the aspect ratio of spheroids in a range
20-25 according to scaling law for isolated spheroidal shape
particle presented on Fig. 1. That means that one could expect
amplification of Raman cross section for molecule placed in the
focus of star-like nanolens to achieve as much as$\;10^{14}$ or
even more. This may open possibility of the single molecule Raman
spectroscopy in the scanning mode in an  AFM type device   with
nanolense placed on the tip of AFM.

The polarization state of the incident electromagnetic field is also of
importance. Qualitatively, the circular polarized wave may be represented as
a sum of linear polarized waves with different directions of the electric
field and appropriate phase shifts. In the symmetric configurations like
those shown in Fig.~\ref{fig2}, each wave excites its own spheroid. The
field distributions depicted in Figs.~\ref{fig4},\ref{fig7} appear as a
result of interference.

\end{document}